\documentclass[twocolumn]{article}

\usepackage{amssymb,amsmath}
\usepackage[]{algorithm2e}
\usepackage{tikz}
\usepackage{subcaption}
\usepackage{graphicx}
\usepackage{authblk}
\usepackage[authoryear]{natbib}

\def\Ttheta{\boldsymbol\theta}
\def\TC{\mathbf C}

\newcommand{\bfB}{{\bf B}}
\newcommand{\bfC}{{\bf C}}
\newcommand{\bfD}{{\bf D}}

\newcommand{\bfK}{{\bf K}}

\newcommand{\bfW}{{\bf W}}

\newcommand{\bfY}{{\bf Y}}

\newcommand{\bftheta}{{\boldsymbol \theta}}

\begin{document}

\title{Neural-networks for geophysicists and their application to seismic data interpretation}

\author[1]{Bas Peters}
\author[1]{Eldad Haber}
\author[2]{Justin Granek}

\affil[1]{University of British Columbia, Vancouver, Canada}
\affil[2]{Computational Geosciences Inc.}
\date{}
\maketitle

\begin{abstract}
Neural-networks have seen a surge of interest for the interpretation of seismic images during the last few years. Network-based learning methods can provide fast and accurate automatic interpretation, provided there are sufficiently many training labels. 
We provide an introduction to the field aimed at geophysicists that are familiar with the framework of forward modeling and inversion. We explain the similarities and differences between deep networks to other geophysical inverse problems and show their utility in solving problems such as lithology interpolation between wells, horizon tracking and segmentation of seismic images. The benefits of our approach are demonstrated on field data from the Sea of Ireland and the North Sea.
\end{abstract}

\section{Introduction}

Deep neural networks (DNNs) have revolutionized computer vision, image processing, and image understanding (see for example \citep{imagenet_cvpr09,krizhevsky2009learning,Ronneberger2015,Goodfellow-et-al-2016} and references within). In particular, deep convolutional networks have solved long standing problems such as image classification, segmentation, debluring, denoising and more. Most of the applications are based on supervised learning, that is, we are given some data and its corresponding interpretation or labels. The goal of the network is to empirically find the connection between the data and its labels.

Seismic interpretation can be viewed as a type of image understanding, where the 3D-image is the seismic cube, and the interpretation of the seismic data, e.g., horizons, faults, etc. are the labeled features that need to be recovered.
Using deep convolution networks is therefore a straight forward extension of existing neural network technology and have been studied recently by many authors (see for example \citep{peters2018multi,peters2019automatic,wu2018deep,doi:10.1190/tle37070529.1,doi:10.1190/1.1484539,Leggett2003,doi:10.1190/segam2018-2998176.1,doi:10.1190/segam2018-2997085.1} and references within).

However, while it seems straight forward to use such algorithms, there are some fundamental differences between vision-related applications to seismic processing.
First, and maybe most importantly is the amount of labeled, or annotated, data available. While in computer vision labeled data is easy to obtain, it is much more difficult to do so for seismic applications. Second, while the labeled data is likely to be correct in vision, it is much more uncertain in seismic interpretation. For example, when viewing an image, it is usually obvious if an object such as a car exists within a frame; on the other hand, two geologists may argue about the existence or the exact location of a particular fault or a deep horizon. This makes the data for the seismic problem biased. Thirdly, even for labeled data, in most applications, the data is not fully labeled and only small portions of it have been annotated. 
Finally, while most vision data is 2D, seismic data is typically in 3D and should therefore be learned in 3D when possible. This makes using Graphical Processing Units (GPUs) challenging due to memory restrictions, especially when the networks are deep and wide.

In this paper, we review and discuss some recent work that we and others have done to tackle some of the challenges when attempting to use deep networks for problems that arise from seismic interpretation. In particular, we address DNNs from a geophysicist's point of view, in terms of network design and optimization. We show that the network can be interpreted as a forward problem while the learning can be interpreted as the inverse problem. Any geophysicist that is familiar with the process of modeling and inversion can therefore understand the process and draw from her previous experiences.
 
In the rest of the paper, we give background information about deep networks. In particular, we discuss the connection between deep networks to differential equations and show that the machine learning problem is similar to other well-studied problems in geophysics such as the full-waveform inversion or electromagnetic forward and inverse problems. This should make it easy for any geophysicist with such background to understand and contribute to the field. We then discuss two different applications that can be tackled using this framework. First, we explain how DNNs can interpolate lithology, given sparse borehole information and seismic data. Next, we show how networks can predict multiple horizons, including branching horizons. We then summarize the paper and discuss and suggest future applications.

\section{Deep Neural Networks - A Geophysicist View}
\label{back}

Supposed we are given data, $\bfD$, and its corresponding label map $\bfC$. If there is a physical basis to obtain $\bfC$ from $\bfD$, then one should use it. For example, assume that $\bfD$ is a velocity model and $\bfC$ is a seismic cube. In this case, one can use the wave equation to obtain $\bfC$ from $\bfD$. However, for many problems in science and engineering such a mapping is unavailable. Since there is no physical basis to recover $\bfC$ from $\bfD$, we turn to an empirical relationship. Many empirical models work well for different applications. For problems where $\bfD$ and $\bfC$ have a spatial interpretation, 
deep neural networks have been successful in capturing the information and generating empirical relationships that hold well in practice.

A deep network is a chain of nonlinear transformations of the data. In particular, we turn to recent work \citep{DBLP:journals/corr/HeZRS15,Chang2017Reversible,HaberRuthotto2017a} that uses residual networks that have the form
\begin{align}
\label{dnn}
&\bfY_{j+1} = \bfY_j - \bfK_j^{\top}\sigma(\bfK_j \bfY_j + \bfB_j), \:\: j=1,\ldots n \\
\nonumber
&\bfY_1 = \bfD
\end{align}
Here, $\bfY_j$ are states, $\bfK_j$ are convolution kernels and $\bfB_j$ are bias vectors. 

Given the network \eqref{dnn} one pushes the data forward through $n$ layers to obtain $\bfY_n$. Given $\bfY_n$ it is possible to predict the label $\bfC$ by simply
multiplying $\bfY_n$ by a matrix $\bfW$. That is
\begin{eqnarray}
\label{class}
\bfC = \bfW \bfY_n
\end{eqnarray}

Let us review the process above from a geophysicist's point of view and show that the above is equivalent to many other forward problems in geophysics. To this end, the deep network \eqref{dnn} can be viewed as a discretization of a physical process, e.g., the wave or Maxwell's equations. From this point of view, $\bfY_j$ are the fields (e.g., acoustic or electromagnetic) and $\bfK_j$ and $\bfB_j$ are model parameters such as seismic velocity or electric conductivity. Just like in any other field, when considering the forward problem we assume that we know the model parameters and therefore we can predict the fields, $\bfY$.
The classification process in Equation~\eqref{class} can be interpreted as projecting the fields to measure some of their properties. A similar process in geophysics is when $\bfW$ is a projection matrix that measures the field at some locations, that is, in receiver positions.

\begin{figure}[!htb]
   \centering
   \includegraphics[width=\columnwidth]{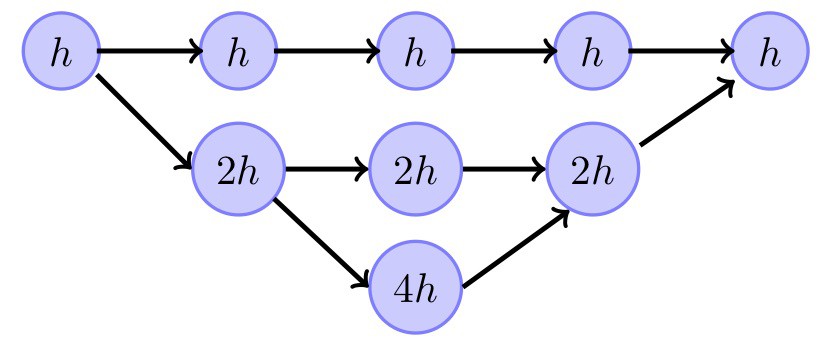}
   \caption{Unet - a number of resnets with scales $h$ (original image), $2h$ (coarsen image) and $4h$. The networks are coupled by restriction and prolongation and are used to deal with data at different resolutions}
   \label{unet.jpg}
\end{figure}

It is important to stress that the network presented in Equation~\eqref{dnn} is just one architecture that we can use. For problems of semantic segmentation it has been shown that coupling a few of these networks, each on a different resolution, gives much better results than using a single resolution. The idea behind such networks is plotted in Figure~\ref{unet.jpg}. We refer the reader to \citep{Ronneberger2015} for more details on efficient network architectures that deal with data with multiple scales.

\bigskip
 
In general, the model parameters $\bfK_j$ and $\bfB_j$ are unknown in practice and need to be calibrated from the data. This process is similar to the process of finding the seismic velocity model or electric conductivity from some measured geophysical data. To this end, we assume that we have some observed labels $\bfC^{\rm obs}$. The learning problem can be framed as a parameter estimation problem, or an inverse problem where we fit the observed labels by minimizing the objective function
\begin{eqnarray}
\label{inv}
\min_{\bftheta} \ \ell(\bfC(\bftheta),\bfC^{\rm obs}) + \alpha R(\bftheta)
\end{eqnarray}
Here we introduce the cumulation of model parameters $\bftheta = \{[\bfK_1,\ldots \bfK_n],[\bfB_1,\ldots \bfB_n]\}$ and a regularization term $R(\bftheta)$. Most literature assumes that $R(\bftheta)$ is a simple Tikhonov regularization or, in the language of deep learning, weight decay, that is
$$ R(\bftheta) = \frac 12 \sum_j \|\bfK_j\|_F^2 + \|\bfB_j\|^2. $$
As we will show next, such basic regularization may not be sufficient for problems that arise from seismic applications, and we review other more appropriate regularization for the problems presented here.

\bigskip

While we have emphasized the similarities between the training problem to other geophysical problems, at this point, it is worthwhile pointing out two fundamental differences between deep learning and geophysical inverse problems. First, and most important, in geophysics we are interested in the model, $\bftheta$. Such a model generally has some physical attributes that we are interested in. 
The model typically represents velocity, conductivity, porosity or other physical properties. 
In machine learning, on the other hand, the model has no real significance. It does not have any physical meaning (that we know of), and therefore it is hard to know what is a ``reasonable'' model. Second, optimizing the objective function in \eqref{inv} is typically done using stochastic gradient descent (SGD) \citep{bottou2008tradeoffs}. It has been shown that using SGD is crucial for the solution of the problem. 

In the following sections, we discuss how we use the setting discussed above to solve a number of practical problems that arise in seismic interpretation.

\section{Applications to seismic interpretation}

In this section, we discuss the application of deep networks to two seismic applications. All applications share the same forward propagation process and the main difference is the way we set up the loss function (misfit) and the regularization.
We find it rather remarkable that similar network architectures work for such different problems, and this emphasizes the strength of deep learning applied to seismic interpretation.

One common feature that most geophysical problems share is that the labels, $\bfC^{\rm obs}$ are not present for the whole seismic image. For example, it is common to have part of the image labeled but not all of it. Another example is that we know only part of a horizon. This is in stark contrast to most computer vision problems where the images are fully labeled. This difference results from the technical difficulty and expertise that is needed to label seismic data. While most non-specialists can identify a cat in an image, an expert may be needed to classify a seismic unit.
However, we note that most applications in geophysics share this type of sparse measurement. For example, we never have a fully observed wave field when considering the full waveform inversion, and the misfit is calculated only on the observable point (where we record the data). We therefore modify common loss functions in DNN training to return the misfit only from the locations where the image is labeled.

\subsection{Interpolation of lithology between wells using seismic data}
\label{lithology}

\begin{figure}[!htb]
   \centering
   \begin{subfigure}[b]{1.0\columnwidth}
   	\centering
   	\includegraphics[width=1.0\columnwidth]{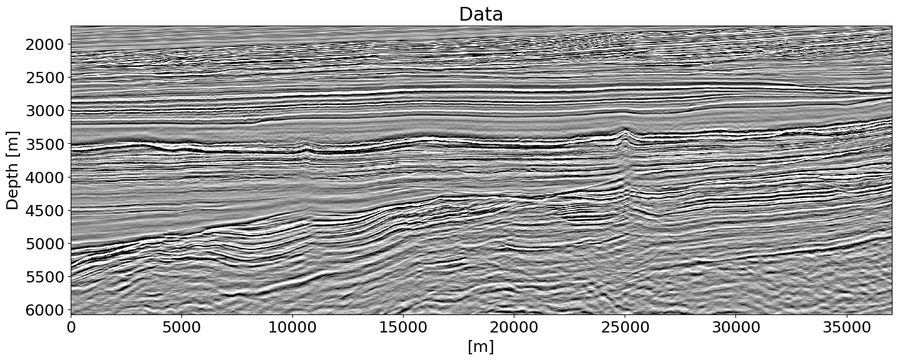}
   	\caption{}
   	\label{/SeismicLithologyWellInterp/withoutReg/data_12}
   \end{subfigure}
   \begin{subfigure}[b]{1.0\columnwidth}
   	\centering
   	\includegraphics[width=1.0\columnwidth]{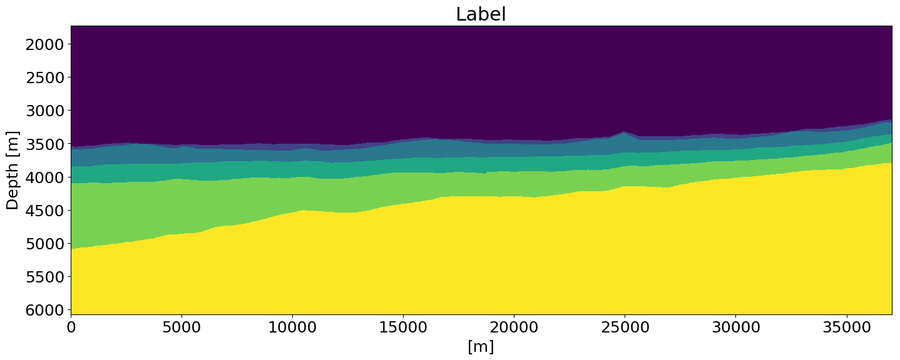}
   	\caption{}
   	\label{/SeismicLithologyWellInterp/withoutReg/label_12}
   \end{subfigure}
      \begin{subfigure}[b]{1.0\columnwidth}
   	\centering
   	\includegraphics[width=1.0\columnwidth]{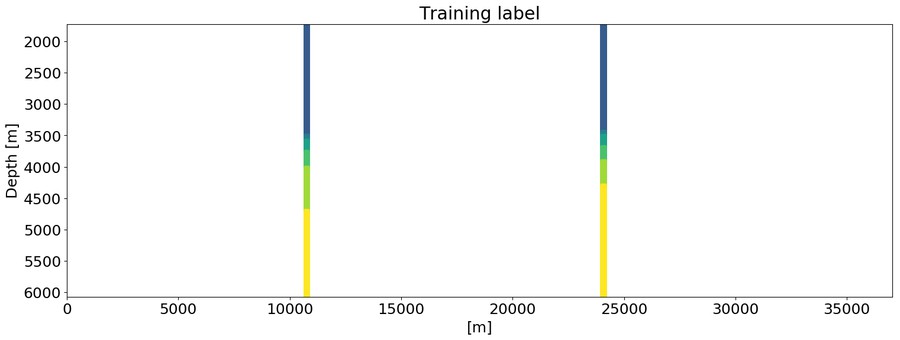}
   	\caption{}
   	\label{/SeismicLithologyWellInterp/withoutReg/label_sub_12}
   \end{subfigure}
   \caption{(a) A slice from a 3D seismic model. This is an example of an input for the network. (b) A fully annotated label image where each color indicates a rock/lithology type of interest. We do not use full labels as the target for our networks, because they are time-consuming to generate. (c) An example of a type of label that we use in our examples. The information corresponds to the lithological units derived from logs in two wells. The white space is not used to measure the misfit or compute a gradient; it is unknown information not used for training the network.}
   \label{test}
 \end{figure}

Consider some boreholes and assume that geological lithology is observed within the boreholes. Our goal is to use lithology information from the wells to interpret the seismic image (Figure \ref{/SeismicLithologyWellInterp/withoutReg/data_12}). 

Specifically, we illustrate the benefits of being able to train on sparse labels such as in Figure \ref{/SeismicLithologyWellInterp/withoutReg/label_sub_12} and predict fully annotated images as in Figure \ref{/SeismicLithologyWellInterp/withoutReg/label_12}. 

When minimizing the loss \eqref{inv} discussed above, artifacts typically appear in the prediction. These artifacts are a result of the lack of data everywhere. To overcome this problem, we propose to add new regularization terms to the loss. This regularization penalizes unwanted oscillations in the prediction maps. 

Note that the true label images that we hope to predict are `blocky'. This implies that the underlying probability of each lithological unit should be smooth. The probability of a particular class changes smoothly from low to high across the interface if the network is well trained.
We propose to mitigate a lack of labels everywhere by using the prior knowledge that the prediction per class should be smooth. This type of prior information fits in the neural-network training process as a penalty function on the output of the network. To this end consider solving an optimization problem of the form
\begin{equation}
\label{partial_ceReg}
L(\bfC(\Ttheta),\TC^{\rm obs})  = - \ell(\bfC(\Ttheta),\TC^{\rm obs}) + \alpha R(\bfY_n(\Ttheta)).
\end{equation}
The regularization $R(\cdot)$ is chosen as
\begin{equation}\label{reg_func}
R(\bfC)  = \frac{1}{2} \| \nabla_h \bfY_n(\bftheta) \|^2
\end{equation}
where $\nabla_h$ is a discrete gradient matrix \citep{haberBook2014} that can be implemented using convolutions with kernels of $\pm1$.

\begin{figure}[!htb]
   \centering
   \begin{subfigure}[b]{1.0\columnwidth}
   	\centering
   	\includegraphics[width=1.0\columnwidth]{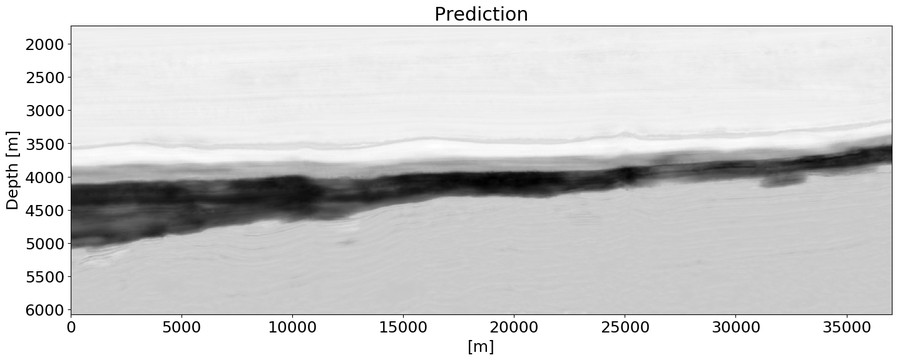}
   	\caption{}
   	\label{/SeismicLithologyWellInterp/withReg/prediction_12}
   \end{subfigure}
   \begin{subfigure}[b]{1.0\columnwidth}
   	\centering
   	\includegraphics[width=1.0\columnwidth]{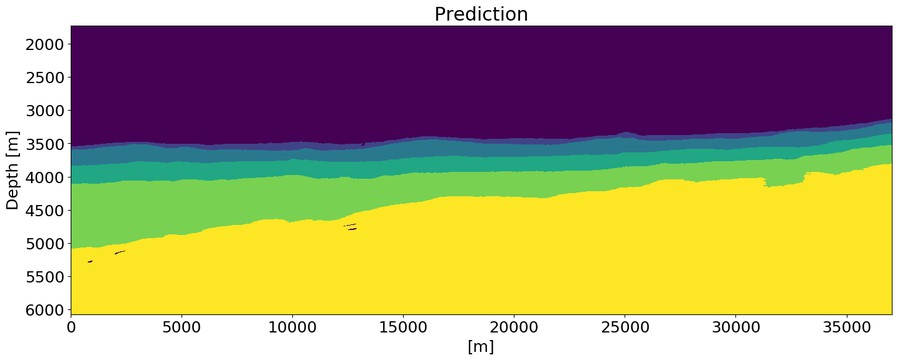}
   	\caption{}
   	\label{/SeismicLithologyWellInterp/withReg/prediction_threshold12}
   \end{subfigure}
   \caption{(a) prediction for a single class and (b) maximum predicted class probability per pixel. Both are the result of training including regularization on the network output.}
\end{figure}

\begin{figure}[!htb]
   \centering
   \includegraphics[width=\columnwidth]{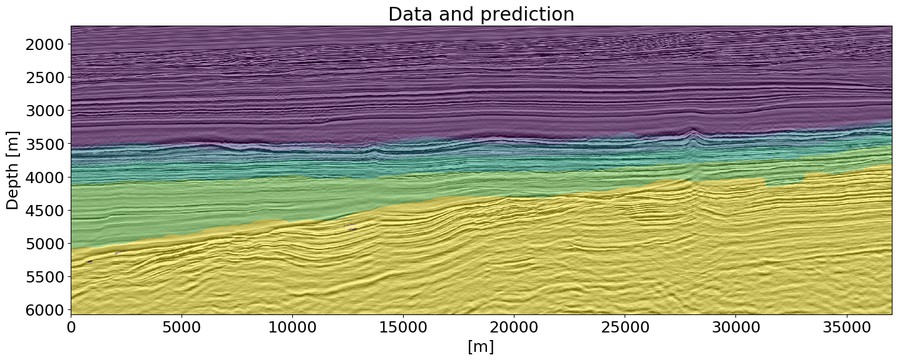}
   \caption{The predicted segmentation from Figure \ref{/SeismicLithologyWellInterp/withReg/prediction_threshold12} (using network output-regularization) overlaid on the seismic input data.}
   \label{/SeismicLithologyWellInterp/withReg/data_pred_thres_12}
\end{figure}

Note that the regularization always applies to the full network output. The output is a full image regardless of sparse sampling of data and/or labels. We can still subsample to introduce randomization or for computational reasons.
The network is trained using the loss function defined in Equation \eqref{partial_ceReg} with quadratic smoothing regularization \eqref{reg_func} applied to the network output. The prediction in Figure \ref{/SeismicLithologyWellInterp/withReg/prediction_12} is smooth and the maximum predicted class probability per pixel in Figure \ref{/SeismicLithologyWellInterp/withReg/prediction_threshold12} is a good approximation to the true map as verified by Figure \ref{/SeismicLithologyWellInterp/withReg/data_pred_thres_12}. Without regularization, the prediction contains many oscillatory artifacts.

\subsection{Horizon tracking by interpolation of scattered picks}
Our second application is tracking a horizon from a small number of horizon picks (seed points) in a few large seismic images. 

Horizon tracking using neural-networks has seen a few time-periods of varying activity \citep{A226265, Veezhinathan1993, doi:10.1190/1.1889749,doi:10.1190/1.1885963, A1543200, doi:10.1190/1.1822449,doi:10.1190/1.1816150}. Algorithms that are not based on learning have also made progress, see, e.g \citep{doi:10.1190/geo2017-0830.1} for recent work that combines and extends multiple concepts on deterministic horizon tracking. 

It was shown previously \citep{peters2018multi} that it is possible to track a single horizon using the U-net based networks and loss-functions that compute losses and gradients based on the sparse labels only. Therefore, there was no need to work in small patches around labeled points or manually generate fully annotated label images. Here we answer two follow-up questions: 1) can we train a network to track more than one horizon simultaneously? 2) How do networks deal with multiple horizons that merge and split? These two questions warrant a new look at the automatic horizon tracking/interpolation problem because results with merging horizons are very rarely published. Especially since there is a renewed surge of interest in using neural networks for seismic interpretation, we need to test the promise of networks against the more challenging situation posed in the above two questions.

\begin{figure}[tbp]
   \centering
   \begin{subfigure}[b]{0.9\columnwidth}
   	\centering
   	\includegraphics[width=1.0\columnwidth]{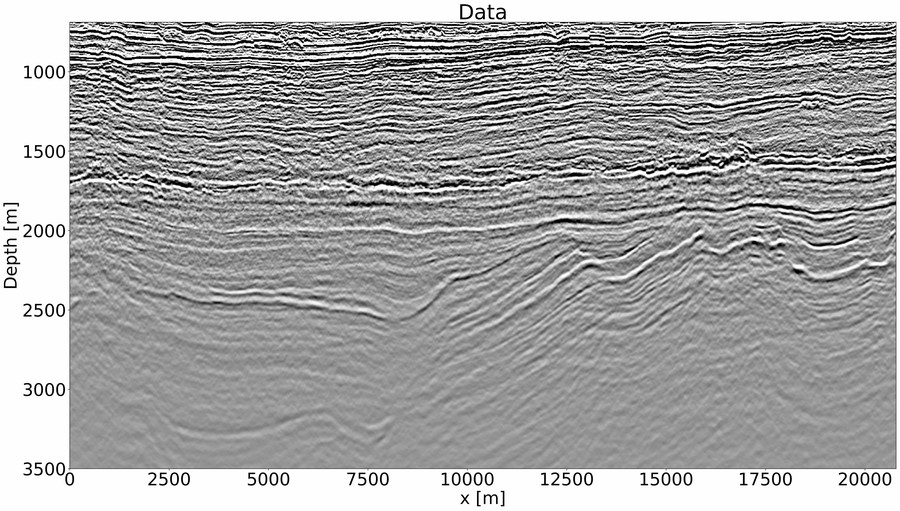}
   	\caption{}
   	\label{/multihorizonfigs/Data_ZX_25}
   \end{subfigure}
   \begin{subfigure}[b]{1.0\columnwidth}
   	\centering
   	\includegraphics[width=0.9\columnwidth]{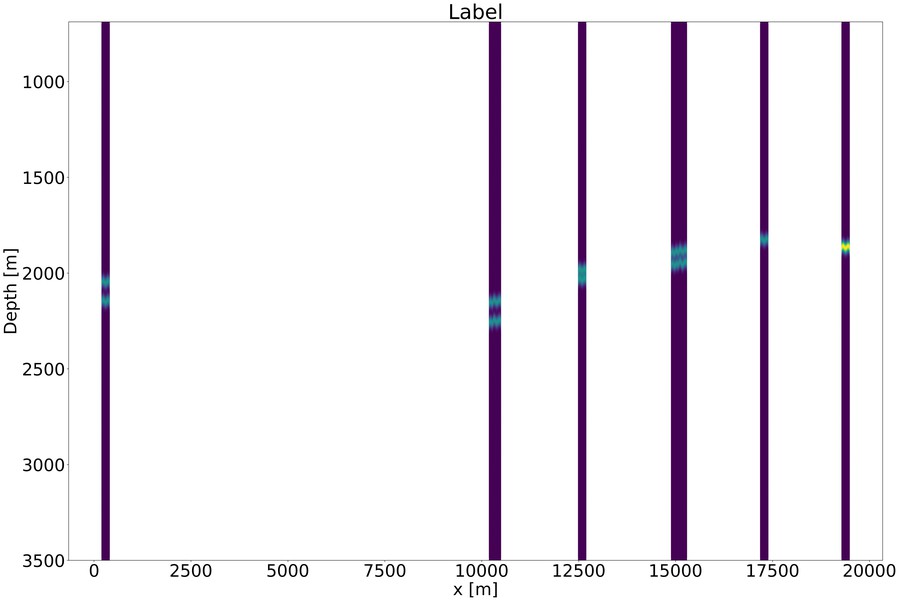}
   	\caption{}
   	\label{/multihorizonfigs/Label_ZX_25}
   \end{subfigure}
   \begin{subfigure}[b]{1.0\columnwidth}
   	\centering
   	\includegraphics[width=0.9\columnwidth]{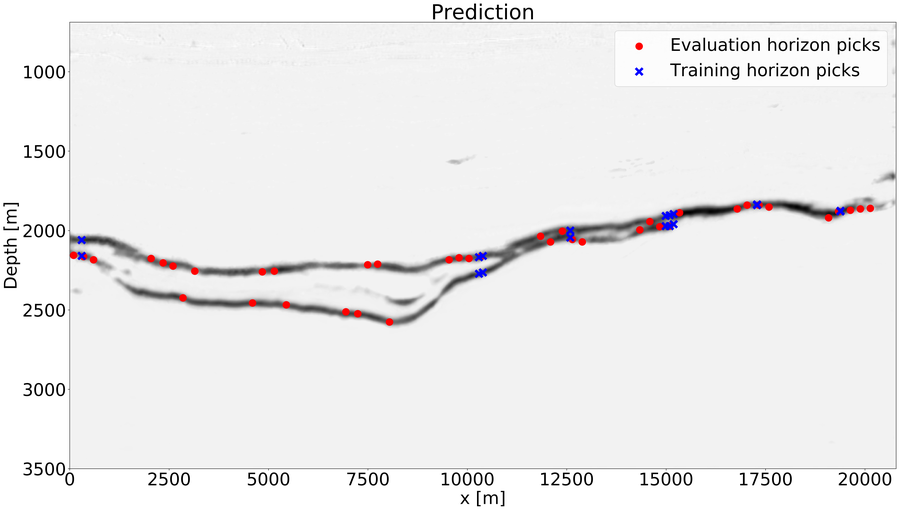}
   	\caption{}
   	\label{/multihorizonfigs/prediction_ZX_25}
   \end{subfigure}
   \begin{subfigure}[b]{1.0\columnwidth}
   	\centering
   	\includegraphics[width=0.9\columnwidth]{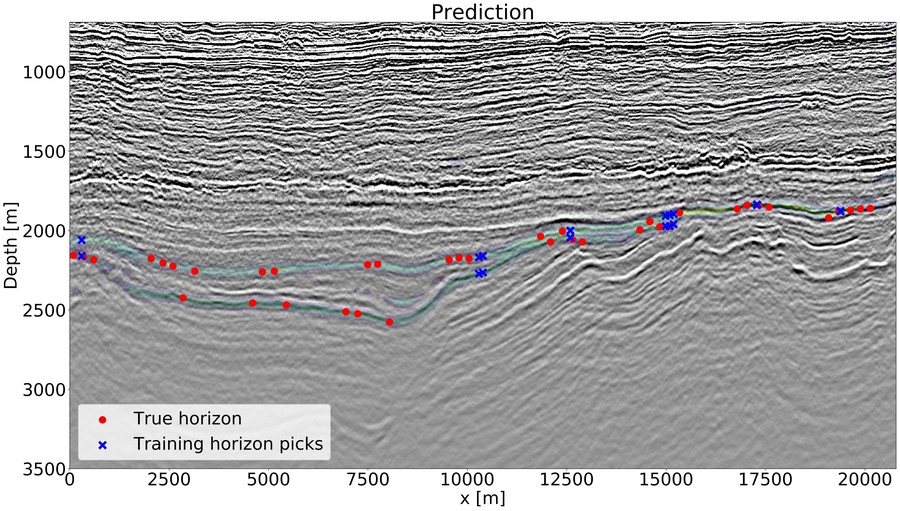}
   	\caption{}
   	\label{/multihorizonfigs/SeisPlusProbs_25}
   \end{subfigure}
   \caption{(a) one of the data images, (b) a label image, about ten columns per image are known, the network never uses the white space. The labels are the convolutions of a Gaussian kernel with the horizon picks. (c) network output with training and testing picks. (d) color-coded network horizon prediction on top of the data.}
\end{figure}

We demonstrate our method using a 3D seismic dataset from the North Sea. One of the $100$ slices is shown in Figure \ref{/multihorizonfigs/Data_ZX_25}. An industrial partner provided us the horizon x-y-z locations, picked by seismic interpreters because their auto-tracking algorithms had difficulties tracking the deeper horizons. We create a label image by convolving the horizon picks (seed points) with a Gaussian kernel in the vertical direction. This procedure adds a sense of uncertainty to the pick. We use approximately $10$ locations per slice for training, as shown in Figure \ref{/multihorizonfigs/Label_ZX_25}. Only the colored columns are used to train the network; in the white space, it is unknown if and where the horizon is. The loss function only uses the information in the known label columns. We see that there are two horizons of interest which merge near the right side of the figure and also get close to each other at the left end. We train a single network to predict both horizons simultaneously, using the non-linear regression and optimization approach detailed in \citep{peters2018multi}. The network design is as described earlier in this work. 

Figure \ref{/multihorizonfigs/prediction_ZX_25} displays the network output, which ideally is the true horizon everywhere convolved with the Gaussian kernel that we used to generate training label images. The training and evaluation picks are plotted on top, and validate that the network is able to predict both horizons accurately, including the point where they merge. In Figure \ref{/multihorizonfigs/SeisPlusProbs_25} we show the network output prediction plotted on top of the seismic data to provide some more insight. The color-coding corresponds to the greyscale intensity of the previous figure. The colors and vertical spread indicate how `sure' the network thinks it is about the prediction.

From the results, we conclude that we can train a single network to simultaneously predict the location of multiple horizons that merge and branch. The symmetric convolutional U-net variant, with the same network architecture as in the previous example, trained by a partial loss-function on a small number of known horizon x-y-z locations achieves excellent results. Data-augmentation and regularization as described in an earlier section can reduce the number of required training x-y-z picks.

\section{Conclusions}

In this paper, we have introduced deep neural networks from an inverse problems point of view. We have shown that the network can be considered as the ``forward problem'' and the training as the ``inverse problem''. We have explored the connection between deep networks to other geophysical inverse problems. We believe that approaching the learning problem in this way allows us to understand better the role of data fitting, regularization, the stability of the network itself, the propagation of noise within the network, and the associated uncertainties; all topics that have received ample treatment in geophysical inverse problems.

We have demonstrated the capability of deep networks to deal with problems that arise from seismic interpretation. In our experience, neural networks can do exceptionally well for such problems given some thought about appropriate regularization and loss or misfit functions.

\bigskip

When solving a particular problem, it is important to realize that geophysical problems are very different from common vision problems. The availability of accurate training data is key to training the network and this can be difficult to obtain in many applications. Another important aspect is the size of the data. While vision problems are typically 2D, many geophysical problems are 3D. We believe that new algorithms should be developed to deal with the size of geophysical images as well as with the uncertainty that is an inherent part of geophysical processing. 

\bibliographystyle{abbrvnat}
\bibliography{seismic_horizon}{}

\begin{thebibliography}{25}
\providecommand{\natexlab}[1]{#1}
\providecommand{\url}[1]{\texttt{#1}}
\expandafter\ifx\csname urlstyle\endcsname\relax
  \providecommand{\doi}[1]{doi: #1}\else
  \providecommand{\doi}{doi: \begingroup \urlstyle{rm}\Url}\fi

\bibitem[Alberts et~al.(2005)Alberts, Warner, and
  Lister]{doi:10.1190/1.1816150}
P.~Alberts, M.~Warner, and D.~Lister.
\newblock Artificial neural networks for simultaneous multi horizon tracking
  across discontinuities.
\newblock In \emph{SEG Technical Program Expanded Abstracts 2000}, pages
  651--653, 2005.
\newblock \doi{10.1190/1.1816150}.
\newblock URL \url{https://library.seg.org/doi/abs/10.1190/1.1816150}.

\bibitem[Bottou and Bousquet(2008)]{bottou2008tradeoffs}
L.~Bottou and O.~Bousquet.
\newblock The tradeoffs of large scale learning.
\newblock In \emph{Advances in neural information processing systems}, pages
  161--168, 2008.

\bibitem[Chang et~al.(2018)Chang, Meng, Haber, Ruthotto, Begert, and
  Holtham]{Chang2017Reversible}
B.~Chang, L.~Meng, E.~Haber, L.~Ruthotto, D.~Begert, and E.~Holtham.
\newblock Reversible architectures for arbitrarily deep residual neural
  networks.
\newblock In \emph{AAAI Conference on AI}, 2018.

\bibitem[Deng et~al.(2009)Deng, Dong, Socher, Li, Li, and
  Fei-Fei]{imagenet_cvpr09}
J.~Deng, W.~Dong, R.~Socher, L.-J. Li, K.~Li, and L.~Fei-Fei.
\newblock {ImageNet: A Large-Scale Hierarchical Image Database}.
\newblock In \emph{CVPR09}, 2009.

\bibitem[Goodfellow et~al.(2016)Goodfellow, Bengio, and
  Courville]{Goodfellow-et-al-2016}
I.~Goodfellow, Y.~Bengio, and A.~Courville.
\newblock \emph{Deep Learning}.
\newblock MIT Press, 2016.

\bibitem[Haber(2014)]{haberBook2014}
E.~Haber.
\newblock \emph{Computational Methods in Geophysical Electromagnetics}.
\newblock SIAM, Philadelphia, 2014.

\bibitem[Haber and Ruthotto(2017)]{HaberRuthotto2017a}
E.~Haber and L.~Ruthotto.
\newblock Stable architectures for deep neural networks.
\newblock \emph{Inverse Problems}, 34\penalty0 (1):\penalty0 014004, dec 2017.
\newblock \doi{10.1088/1361-6420/aa9a90}.

\bibitem[Harrigan et~al.(1992)Harrigan, Kroh, Sandham, and Durrani]{A226265}
E.~Harrigan, J.~R. Kroh, W.~A. Sandham, and T.~S. Durrani.
\newblock Seismic horizon picking using an artificial neural network.
\newblock In \emph{[Proceedings] ICASSP-92: 1992 IEEE International Conference
  on Acoustics, Speech, and Signal Processing}, volume~3, pages 105--108 vol.3,
  March 1992.
\newblock \doi{10.1109/ICASSP.1992.226265}.

\bibitem[He et~al.(2015)He, Zhang, Ren, and Sun]{DBLP:journals/corr/HeZRS15}
K.~He, X.~Zhang, S.~Ren, and J.~Sun.
\newblock Deep residual learning for image recognition.
\newblock \emph{CoRR}, abs/1512.03385, 2015.
\newblock URL \url{http://arxiv.org/abs/1512.03385}.

\bibitem[Huang et~al.(2005)Huang, Chang, Hsieh, Hsieh, Wang, and
  Tsai]{A1543200}
K.-Y. Huang, C.-H. Chang, W.-S. Hsieh, S.-C. Hsieh, L.~K. Wang, and F.-J. Tsai.
\newblock Cellular neural network for seismic horizon picking.
\newblock In \emph{2005 9th International Workshop on Cellular Neural Networks
  and Their Applications}, pages 219--222, May 2005.
\newblock \doi{10.1109/CNNA.2005.1543200}.

\bibitem[Huang(2005)]{doi:10.1190/1.1885963}
K.~â. Huang.
\newblock Hopfield neural network for seismic horizon picking.
\newblock In \emph{SEG Technical Program Expanded Abstracts 1997}, pages
  562--565, 2005.
\newblock \doi{10.1190/1.1885963}.
\newblock URL \url{https://library.seg.org/doi/abs/10.1190/1.1885963}.

\bibitem[Krizhevsky and Hinton(2009)]{krizhevsky2009learning}
A.~Krizhevsky and G.~Hinton.
\newblock Learning multiple layers of features from tiny images.
\newblock Technical report, Citeseer, 2009.

\bibitem[Kusuma and Fish(2005)]{doi:10.1190/1.1822449}
T.~Kusuma and B.~C. Fish.
\newblock Toward more robust neural‐network first break and horizon pickers.
\newblock In \emph{SEG Technical Program Expanded Abstracts 1993}, pages
  238--241, 2005.
\newblock \doi{10.1190/1.1822449}.
\newblock URL \url{https://library.seg.org/doi/abs/10.1190/1.1822449}.

\bibitem[Leggett et~al.(2003)Leggett, Sandham, and Durrani]{Leggett2003}
M.~Leggett, W.~A. Sandham, and T.~S. Durrani.
\newblock \emph{Automated 3-D Horizon Tracking and Seismic Classification Using
  Artificial Neural Networks}, pages 31--44.
\newblock Springer Netherlands, Dordrecht, 2003.
\newblock ISBN 978-94-017-0271-3.
\newblock \doi{10.1007/978-94-017-0271-33}.
\newblock URL \url{https://doi.org/10.1007/978-94-017-0271-33}.

\bibitem[Liu et~al.(2005)Liu, Xue, and Li]{doi:10.1190/1.1889749}
X.~Liu, P.~Xue, and Y.~Li.
\newblock Neural network method for tracing seismic events.
\newblock In \emph{SEG Technical Program Expanded Abstracts 1989}, pages
  716--718, 2005.
\newblock \doi{10.1190/1.1889749}.
\newblock URL \url{https://library.seg.org/doi/abs/10.1190/1.1889749}.

\bibitem[Lowell and Paton(2018)]{doi:10.1190/segam2018-2998176.1}
J.~Lowell and G.~Paton.
\newblock Application of deep learning for seismic horizon interpretation.
\newblock In \emph{SEG Technical Program Expanded Abstracts 2018}, pages
  1976--1980, 2018.
\newblock \doi{10.1190/segam2018-2998176.1}.
\newblock URL
  \url{https://library.seg.org/doi/abs/10.1190/segam2018-2998176.1}.

\bibitem[Peters et~al.(2018)Peters, Granek, and Haber]{peters2018multi}
B.~Peters, J.~Granek, and E.~Haber.
\newblock Multi-resolution neural networks for tracking seismic horizons from
  few training images.
\newblock \emph{arXiv preprint arXiv:1812.11092}, 2018.

\bibitem[Peters et~al.(2019)Peters, Granek, and Haber]{peters2019automatic}
B.~Peters, J.~Granek, and E.~Haber.
\newblock Automatic classification of geologic units in seismic images using
  partially interpreted examples.
\newblock \emph{arXiv preprint arXiv:1901.03786}, 2019.

\bibitem[Poulton(2002)]{doi:10.1190/1.1484539}
M.~M. Poulton.
\newblock Neural networks as an intelligence amplification tool: A review of
  applications.
\newblock \emph{GEOPHYSICS}, 67\penalty0 (3):\penalty0 979--993, 2002.
\newblock \doi{10.1190/1.1484539}.
\newblock URL \url{https://doi.org/10.1190/1.1484539}.

\bibitem[Ronneberger et~al.(2015)Ronneberger, Fischer, and
  Brox]{Ronneberger2015}
O.~Ronneberger, P.~Fischer, and T.~Brox.
\newblock U-net: Convolutional networks for biomedical image segmentation.
\newblock \emph{Medical Image Computing and Computer-Assisted Intervention –
  MICCAI 2015}, page 234–241, 2015.
\newblock ISSN 1611-3349.
\newblock \doi{10.1007/978-3-319-24574-428}.
\newblock URL \url{http://dx.doi.org/10.1007/978-3-319-24574-428}.

\bibitem[Veezhinathan et~al.(1993)Veezhinathan, Kemp, and
  Threet]{Veezhinathan1993}
J.~Veezhinathan, F.~Kemp, and J.~Threet.
\newblock A hybrid of neural net and branch and bound techniques for seismic
  horizon tracking.
\newblock In \emph{Proceedings of the 1993 ACM/SIGAPP Symposium on Applied
  Computing: States of the Art and Practice}, SAC '93, pages 173--178, New
  York, NY, USA, 1993. ACM.
\newblock ISBN 0-89791-567-4.
\newblock \doi{10.1145/162754.162863}.
\newblock URL \url{http://doi.acm.org/10.1145/162754.162863}.

\bibitem[Waldeland et~al.(2018)Waldeland, Jensen, Gelius, and
  Solberg]{doi:10.1190/tle37070529.1}
A.~U. Waldeland, A.~C. Jensen, L.-J. Gelius, and A.~H.~S. Solberg.
\newblock Convolutional neural networks for automated seismic interpretation.
\newblock \emph{The Leading Edge}, 37\penalty0 (7):\penalty0 529--537, 2018.
\newblock \doi{10.1190/tle37070529.1}.
\newblock URL \url{https://doi.org/10.1190/tle37070529.1}.

\bibitem[Wu and Zhang(2018)]{wu2018deep}
H.~Wu and B.~Zhang.
\newblock A deep convolutional encoder-decoder neural network in assisting
  seismic horizon tracking.
\newblock \emph{arXiv preprint arXiv:1804.06814}, 2018.

\bibitem[Wu and Fomel(2018)]{doi:10.1190/geo2017-0830.1}
X.~Wu and S.~Fomel.
\newblock Least-squares horizons with local slopes and multigrid correlations.
\newblock \emph{GEOPHYSICS}, 83\penalty0 (4):\penalty0 IM29--IM40, 2018.
\newblock \doi{10.1190/geo2017-0830.1}.
\newblock URL \url{https://doi.org/10.1190/geo2017-0830.1}.

\bibitem[Zhao(2018)]{doi:10.1190/segam2018-2997085.1}
T.~Zhao.
\newblock Seismic facies classification using different deep convolutional
  neural networks.
\newblock In \emph{SEG Technical Program Expanded Abstracts 2018}, pages
  2046--2050, 2018.
\newblock \doi{10.1190/segam2018-2997085.1}.
\newblock URL
  \url{https://library.seg.org/doi/abs/10.1190/segam2018-2997085.1}.

\end{thebibliography}
\end{document}